\documentclass[sn-mathphys-ay,iicol,pdflatex]{sn-jnl}

\usepackage{graphicx}%
\usepackage{multirow}%
\usepackage{amsmath,amssymb,amsfonts}%
\usepackage{amsthm}%
\usepackage{mathrsfs}%
\usepackage[title]{appendix}%
\usepackage{xcolor}%
\usepackage{textcomp}%
\usepackage{manyfoot}%
\usepackage{booktabs}%
\usepackage{algorithm}%
\usepackage{algorithmicx}%
\usepackage{algpseudocodex}%
\usepackage{listings}%
\usepackage{siunitx}
\usepackage{pdflscape}
\usepackage{braket}
\usepackage{csquotes}
\usepackage{cleveref}
\usepackage{setspace}
\usepackage{caption}
\usepackage{subcaption}

\usepackage{etoolbox}           
\newcommand{\B}{\fontseries{b}\selectfont} 
\robustify\B

\makeatletter
\expandafter\patchcmd\csname\string\algorithmic\endcsname{\itemsep\z@}{\itemsep=0.2ex}{}{}
\makeatother

\begin{document}

\title{A hybrid learning agent for episodic learning tasks with unknown target distance}

\author*[1]{\fnm{Oliver} \sur{Sefrin}}\email{oliver.sefrin@dlr.de}
\author[1]{\fnm{Sabine} \sur{W\"olk}\nomail}

\affil[1]{\orgdiv{Institute of Quantum Technologies}, \orgname{German Aerospace Center (DLR)}, \orgaddress{\street{Wilhelm-Runge-Straße 10}, \city{Ulm}, \postcode{89081}, \country{Germany}}}

\abstract{
     The ``hybrid agent for quantum-accessible reinforcement learning'', as defined in~\citep{hamann2022performance}, provides a proven quasi-quadratic speedup and is experimentally tested. 
     However, the standard version can only be applied to episodic learning tasks with fixed episode length. 
     In many real-world applications, the information about the necessary number of steps within an episode to reach a defined target is not available in advance and especially before reaching the target for the first time. 
     Furthermore, in such scenarios, classical agents have the advantage of observing at which step they reach the target. 
     Whether the hybrid agent can provide an advantage in such learning scenarios was unknown so far.
     
     In this work, we introduce a hybrid agent with a stochastic episode length selection strategy to alleviate the need for knowledge about the necessary episode length.
     Through simulations, we test the adapted hybrid agent's performance versus classical counterparts.
     We find that the hybrid agent learns faster than corresponding classical learning agents in certain scenarios with unknown target distance and without fixed episode length. 
}


\maketitle

\section{Introduction}
\label{sec:intro}

Reinforcement Learning~(RL), the Machine Learning framework related to learning through interaction and experience, has shown tremendous success in recent years, surpassing human capability in Atari games~\citep{mnih2015human} or the board game Go~\citep{silver2017mastering}, amongst many others.
One of the main reasons for its success is the ever-growing computational power of classical hardware, which allows the implementation of Deep Learning techniques in RL such as Deep Q-Learning~(DQN)~\citep{mnih2015human}.

However, many problems and problem classes still prove to be difficult even to modern supercomputers due to their unfavorable scaling with the problem size.
Here, quantum computation~\citep{nielsen2002quantum} with the prospect of polynomial or even exponential advantages in problem complexity has excited researchers across many disciplines and even created new research fields.
Among the latter, Quantum Machine Learning~(QML)~\citep{biamonte2017quantum} has emerged with the idea of combining the computational benefits of quantum computation with the proven effectiveness of machine learning approaches.
Given the current state of quantum computing hardware in the so-called noisy intermediate-scale quantum (NISQ) era~\citep{preskill2018nisq}, research in QML has focused on algorithms of low circuit depth and width, such as variational algorithms~\citep{cerezo2021variational}, or even so-called quantum-inspired methods such as tensor networks~\citep{biamonte2017tensor,orus2014practical,bridgeman2017hand,huggins2019towards}.
Whereas these methods have the benefit of being applicable on current quantum devices (or even on classical machines in the case of quantum-inspired ansätze), their actual advantage compared to classical methods remains unclear~\citep{schuld2023quantum,cerezo2024does}.

An example for a QML algorithm with a provable speedup which is, however, not NISQ-ready, is the so-called ``hybrid agent for quantum-accessible reinforcement learning''~\citep{hamann2022performance}. 
Here, a learning agent learns to solve a given problem by interacting via a set of actions with a problem environment.
Based on amplitude amplification~\citep{grover1997quantum,brassard2002quantum},  a quasi-quadratic speedup in terms of the sample complexity compared to corresponding classical agents was proven for a class of RL environments called \textit{deterministic and strictly episodic (DSE)} environments.
These environments are reset to an initial state after a fixed number of interaction steps defining the \textit{episode length}.
Further, choosing an action in the current state yields one subsequent state with certainty.
The \textit{Gridworld} or maze scenario with a fixed episode length is one example of a DSE environment and serves as a toy model in this work.

\begin{figure}[t]
    \centering
    \includegraphics[width=\columnwidth]{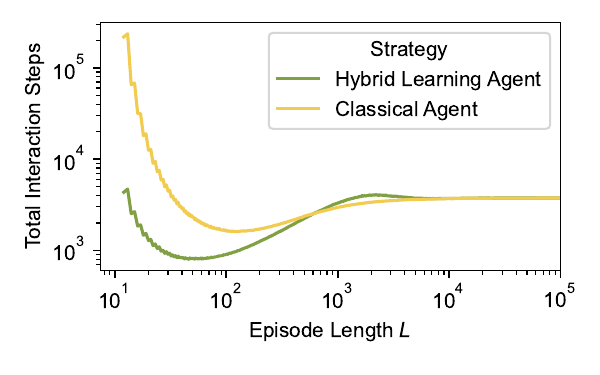}
    \caption{Comparison of the expected number of total interaction steps performed before finding a defined target for the first time depending on the episode length~$L$.
    The results stem from a simulation with a Gridworld of base size~$7\times 7 $ and an outer wall distance of 16 (see~\Cref{sec:methods:sub:simulation} for information about the Gridworld layouts).
    }
    \label{fig:expected_cost_fix_length}
\end{figure}

While many learning scenarios are indeed episodic, a large subset of these is not strictly episodic.
That is, episodes may differ in length because, e.g., their end is coupled to reaching some rewarded target state such that the episode length depends on the brevity of the solution found.
The Gridworld environment without a fixed episode length is an example for such a learning scenario.
In general, the hybrid agent~\citep{hamann2022performance} can easily be applied to a deterministic and non-strictly episodic environment by choosing a fixed episode length $L$ after which the episode ends, whether the target was reached or not. 
Yet, the chosen episode length $L$ has crucial influence on the hardness of the learning problem and the expected total number of interaction steps to learn to solve the problem, both in the classical and the hybrid case.
Fig.~\ref{fig:expected_cost_fix_length} illustrates this influence. 
Here, we show the expected number of interaction steps, summed over all episodes, until the defined goal was reached for the first time for a classical and a hybrid agent depending on the fixed episode length $L$ for a given maze example. 
On the one hand, choosing $L$ small renders the problem hard or even unsolvable for an untrained agent. 
On the other hand, choosing a large $L$ can lead to a high probability $p_\text{init}(L)$ to achieve the defined target within one episode even for an untrained agent. 
This reduces the number of necessary episodes but makes a large number of interaction steps per episode necessary.
For small episode lengths, corresponding to small success probabilities~$p_\mathrm{init}(L)$, the hybrid agent offers a quasi-quadratic speedup compared to the classical agent.
For intermediate episode lengths, the hybrid agent's quantum overhead results in a slightly worse performance, before it finally converges to the classical agent's behavior and thus to its performance in the limit of large episode lengths and $p_\mathrm{init}(L)\rightarrow 1$.
(For a detailed discussion, see Appendix~\ref{app:fixed-episode-length}.)

In summary, the choice of $L$ can have a larger influence on the necessary effort to reach the goal than whether we use a hybrid or a purely classical agent.
If enough information about the learning problem were available, an optimal episode length could be inferred.
However, this is usually not the case. 
Furthermore, a classical learning agent can observe when it reaches the target and then end the episode. 
In quantum mechanics, observation usually suppresses the effect of interference, which is necessary to gain a quantum advantage. 
Thus, a predefined episode length is necessary for the hybrid agent. 
Whether the hybrid agent can outperform  classical agents which use the advantage of a flexible episode length and if so, in which scenarios, has not been investigated so far.

In this work, we introduce a hybrid agent  with a flexible episode length selection strategy for the case of a deterministic and episodic environment with an unknown target distance.
More precisely, we introduce a probabilistic condition which doubles the current episode length when triggered until the target is found for the first time. 
Reaching the target for the first time is especially challenging while extremely important.
Indeed, after having reached the target, the length of the found sequence of actions is obviously sufficient and serves as an upper bound for the optimal episode length.
The necessity of reaching the target for the first time efficiently is further underlined by the fact that, at the start of learning, the RL algorithm's success probabilities are usually the lowest.
This indicates that the search for reaching the target for the first time in such a sparse reward environment takes up a significant amount of the total learning time.

By interweaving our probabilistic episode length adaptation with the underlying amplitude amplification algorithm, we manage to solve learning problems without fixed episode length with no additional hyperparameters introduced.
We test the extended hybrid agent versus a corresponding classical agent with the same episode length selection strategy and a classical agent with flexible episode length. 
This latter agent has no fixed episode length and ends an episode only when finding the target. 
By investigating various layouts of the Gridworld problem, and comparing the performance of the different agents, we can distinguish different learning scenarios where either our hybrid agent or a classical agent with a completely flexible episode length is favorable.

The article is structured as follows:
in \Cref{sec:background}, the hybrid agent is introduced in more detail as well as placed in the broader context of Quantum Reinforcement Learning (QRL).
Next, we present the adapted algorithm and explain the simulation methodology in \Cref{sec:methods}.
In \Cref{sec:results}, we define a new figure of merit tailored to the new problem setting and motivate it before presenting the results. 
Finally, we provide a conclusion and an outlook in \Cref{sec:conclusion}.

\section{Background}
\label{sec:background}

\subsection{Reinforcement Learning}
\label{sec:background:sub:rl}

In RL~\citep{sutton2018reinforcement}, an agent interacts with an environment in a sequence of discrete time steps $t\in\mathbb{N}_0$.
The interaction starts with an initial percept, or state, $s_0\in \mathcal{S}$ from the set of possible percepts $\mathcal{S}$, which is given to the agent by the environment.
At each time step $t\geq 1$, the agent selects an action $a_t$ from the set of possible actions $\mathcal{A}$ based on the previous percept $s_{t-1}$.
This action selection is governed by the agent's current \textit{policy} $\pi(a|s)$, which is a probability distribution over the set of actions conditional to the current percept.
Subsequently, the environment responds with the next percept $s_t$ as well as a real-valued reward $r_t$.
Generally, the response of the environment is probabilistic, with probability distribution 
\begin{align}\label{eq:mdp-dymanics}
    \tau(s_t, r_t|s_{t-1}, a_t).
\end{align}
Since the probability function only depends on the previous interaction step, it fulfills the Markov property. 
Hence, this type of RL interaction is a so-called finite \textit{Markov Decision Process} (MDP).

In \textit{deterministic} environments such as Gridworld, \cref{eq:mdp-dymanics} is a trivial probability distribution in the sense that it returns unity for one specific combination of percept and reward value $(s_t, r_t)$ and zero for any other combination.
To simplify the notation in \Cref{alg:prob-jump}, we can thus introduce functions $S: \mathcal{S} \times \mathcal{A} \rightarrow \mathcal{S}$ and $R: \mathcal{S} \times \mathcal{A} \rightarrow \mathbb{R}$ which return the next, deterministic percept $s_t$ and reward $r_t$, respectively:
\begin{align}
    r_t &= R(s_{t-1}, a_t) \\
    s_t &= S(s_{t-1}, a_t)
\end{align}
The agent's task is to adapt its policy such as to maximize the expectation value of the \textit{cumulative reward}
\begin{align}
    \mathbb{E}_\pi\left[ \sum\limits_{t=1}^\infty \gamma^t r_t \right]\,,
\end{align}
where the subscript $\pi$ indicates the expectation value upon following policy $\pi$.
The coefficient $\gamma\in(0,1]$ is the so-called discount value, which adjusts a trade-off between current and future rewards.

\subsection{Hybrid Learning Agent}
\label{sec:background:sub:hybrid-agent}

Our approach builds on the QRL algorithm coined ``hybrid agent for quantum-accessible reinforcement learning''~\citep{hamann2022performance}.
This algorithm is embedded in a quantum communication scenario where the RL agent and the DSE environment interact by exchanging quantum states. 
That is, each action from the set of allowed actions $a \in \mathcal{A}$ is mapped to a quantum state $\ket{a}_A$ and the set of states $\{ \,\ket{a} \mid a\in\mathcal{A} \, \}$ forms an orthonormal basis. 
Similarly, the set of percepts $\{ \, s \mid s \in \mathcal{S} \, \}$ is mapped to orthonormal states $\{ \ket{s}_S \}$, and the set of rewards $\{ \,r \mid r\in \mathcal{R}\, \}$ to orthonormal states $\{ \ket{r}_R\}$.
Here, the indices $A$, $S$, and $R$ indicate the Hilbert spaces for the quantized actions~$\mathcal{H}_A$, percepts~$\mathcal{H}_S$, and rewards~$\mathcal{H}_R$, respectively.
We map the initial percept $s_0$ to 
$$\ket{\vec{s}_\mathrm{init}}_S = \ket{s_0,\varnothing, \dots,\varnothing}_S \in \mathcal{H}_S^{\otimes(L+1)},$$ 
with a single-percept default state $\ket{\varnothing}_S \in \mathcal{H}_S$, as the initial quantum state for the sequence of percepts and use $\ket{\vec{0}}_R \in \mathcal{H}_R^{\otimes L}$ as the initial quantum state for the sequence of rewards.
Then, in the quantum communication scenario, the response of the environment to the sequence of actions $\vec{a}=(a_1,\dots,a_L)$ within one episode of $L$ interaction steps can be modeled as a unitary $U_\mathrm{env}$ acting on the multipartite state $\ket{\psi}=\ket{\vec{a}}_A \ket{\vec{s}_\mathrm{init}}_S \ket{\vec{0}}_R$ such that 
$$U_\mathrm{env}\ket{\psi}= \ket{\vec{a}}_A \ket{\vec{s}}_S \ket{\vec{r}}_R.$$
In our RL scenario, we assume a single binary reward $r$ is given at the end of each episode to simplify the learning scenario, such that the reward Hilbert space is two-dimensional.

With $\alpha_k$\footnote{$\alpha_k = 1$ for basic environments with action-independent percepts $s$ and $\alpha_k=2$ for environments in which intermediate percepts $s$ are action-dependent~\citep{hamann2022performance}.\label{ftn:alpha-k}} queries of the environment unitary $U_\mathrm{env}$, it was shown that an effective phase-kickback oracle 
\begin{align}\label{eq:phase-kickback}
    O_E \ket{\vec{a}}_A \ket{\vec{s}_\mathrm{init}}_S \ket{-}_R = (-1)^{r(\Vec{a})} \ket{\vec{a}}_A \ket{\vec{s}_\mathrm{init}}_S \ket{-}_R
\end{align}
can be realized~\citep{dunjko2016quantum} when initializing the reward register in the state $\ket{-}_R= \frac{1}{\sqrt{2}}(\ket{0}_R-\ket{1}_R)$. 
Based on this oracle, a hybrid quantum classical learning agent has been defined in~\citep{saggio2021experimental,hamann2022performance}.

This hybrid learning agent consists of two parts: a quantum part and a classical part.
In the quantum part of the hybrid learning agent~\citep{hamann2022performance}, the agent prepares instead of a single sequence of actions~$\ket{\vec{a}}$ a superposition of possible action sequences $\sum_{\vec{a}}c_{\vec{a}}\ket{\vec{a}}$ where $|c_{\vec{a}}|^2$~is equal to the probability that the agent chooses the action sequence $\vec{a}$ according to its current policy~$\pi$. 
By interacting with the environment, the agent applies a Grover operator on $\sum_{\vec{a}}c_{\vec{a}}\ket{\vec{a}}$ based on the oracle $O_E$. 
With this Grover operator, the amplitudes $\lbrace\, c_{\vec{a}} \,| \,r(\vec{a}) > 0 \rbrace$ of rewarded action sequences can be amplified which enables a quadratic speedup in query complexity~\citep{grover1997quantum,brassard2002quantum}. 
Following the quantum part of the algorithm, the classical part of the agent starts by measuring the action register. 
Consecutively, one classical episode with the measured action sequence $\vec{a}$ is performed to determine the corresponding sequence of percepts $\vec{s}\left(\vec{a}\right)$ and the corresponding reward $r\left(\vec{a}\right)$ for the measured sequence of actions.
This is necessary to infer the actual sequence of percepts encountered by the agent as well as the reward information, since this information was uncomputed in the amplitude amplification procedure to allow for interference. 
Finally, the policy~$\pi$ of the agent can be updated according to some chosen update rules such as Q-Learning~\citep{watkins1992q} or Projective Simulation~\citep{briegel2012projective}. 
Then, the agent proceeds by starting the next quantum part until a predefined ending condition~\citep{hamann2022performance} is met.

The speedup of the hybrid learning agent has been verified in a proof-of-principle experiment using a nanophotonic processor~\citep{saggio2021experimental}.
Likewise, a speedup in decision-making using a quantum walk search approach has been formulated~\citep{paparo2014quantum} and experimentally verified~\citep{sriarunothai2018speeding} for a variant of the Projective Simulation algorithm called ``Reflecting Projective Simulation''~\citep{briegel2012projective}.
A first extension of the standard RL scenario for the hybrid learning agent concerning changing oracles was investigated in~\citep{hamann2021quantum}.

\subsection{Related Work}
\label{sec:background:sub:related-work}

QRL~\citep{dong2008quantum}, just as Quantum Machine Learning, can be interpreted in many different ways, according to which we aim to structure this overview of related work.

In the broader sense, QRL can be understood as the application of classical RL~\citep{sutton2018reinforcement} to problems related to quantum computing or quantum technologies.
This comprises the usage of RL for quantum circuit optimization~\citep{ostaszewski2021reinforcement,lockwood2022optimizing,fosel2021quantum,ruiz2024quantum,rapp2024reinforcement}, quantum control~\citep{sivak2022model,fosel2018reinforcement,bukov2018rl-quantum-control,guatto2024improving}, or quantum error correction~\citep{nautrup2019optimizing,sivak2023real}.

Another well-established category is QRL with parametrized quantum circuits (PQCs)~\citep{jerbi2021parametrized}.
Here, PQCs encode the RL agent's current policy (in policy gradient algorithms such as PPO~\citep{schulman2017proximal} or DPG/DDPG~\citep{silver2014dpg,lillicrap2019continuous}) or encode an approximate action-value function such as in DQN~\citep{mnih2015human}.
In a hybrid approach, the PQC's parameters are updated using a classical feedback loop.
Using simulated or real quantum hardware, these methods have been applied to standard Gymnasium~\citep{kwiatkowski2024gymnasium} (previously OpenAI Gym~\citep{brockman2016openai}) environments such as Cart Pole, Frozen Lake, or Atari Games~\citep{jerbi2021parametrized,skolik2022quantum,chen2020variational,lockwood2020reinforcement,lockwood2021playing} as well as maze problems~\citep{hohenfeld2024quantum,chen2024deep}.

More recent approaches aim at speeding up the learning process using quantum sub-routines~\citep{ganguly2023quantum,zhong2024provably}, finding better policies with a combined approach of using quantum phase estimation~\citep{kitaev1995quantum} and Grover's search algorithm~\citep{grover1997quantum,wiedemann2023quantum}, or combining quantum computing with the policy iteration algorithm~\citep{cherrat2023quantum}.

A different approach to the maze or Gridworld problem is shown in~\citep{pozza2022quantum}.
Here, a classical RL agent is trained to modify the maze's walls such as to maximize the escape probability with a quantum random walk in a given time interval.

\section{Methods}
\label{sec:methods}

\subsection{Algorithm}
\label{sec:methods:sub:algo}

Our hybrid learning agent uses, similar to~\citep{hamann2022performance}, the variation of Grover's algorithm where the number of solutions and thus the optimal number $k_\text{opt}$ of amplitude amplification (AA) iterations is unknown~\citep{boyer1998tight}.
This variation of the Grover algorithm appears also slightly varied as \texttt{QSearch} in~\citep{brassard2002quantum} and is necessary due to the fact that in a RL problem, the \textit{success probability} of being rewarded is typically unknown.

The main ingredient of what we call from here on \textit{Boyer's algorithm} is a flexible interval $[0,m)$ with $m\in \mathbb{R}^+$, from which the integer number of AA iterations $k$ is uniformly sampled.
Starting from $m=1$, the interval upper bound is multiplied by a constant factor $\lambda \in (1, 2)$ each time that the measurement yields no success.
Once the parameter $m$ reaches $\sqrt{1/p_\mathrm{min}}$, with $p_\mathrm{min}$ being a lower bound for the current success probability, it is not increased further.
At this point, Boyer's algorithm reaches its so-called \textit{critical stage}, at which the success probability in each AA round is known to be at least $1/4$, supposed that a rewarded item exists~\citep{boyer1998tight}.

\begin{algorithm}[t]
\caption{Probabilistic Hybrid Algorithm}\label{alg:prob-jump}
\begin{algorithmic}[1]
    \vspace{3pt}
    \Require initial percept $s_0$, set of actions $\mathcal{A}$, policy $\pi$, percept response function $S$, reward function $R$, Grover operator $G$
    \State $L \gets 1$ 
    \State $m \gets 1$
    \State $\lambda \gets 5/4$ 
    \State $N_\mathrm{act}=0$ \Comment{step counter}
    \While{\texttt{true}}
        \LComment{probabilistic doubling of $L$}
        \State $q \gets \mathrm{random \;number \;in \;[0,1]}$
        \If{$q < \frac{2\log(m)}{L \log(| \mathcal{A} |)}$}
            \State $L \gets 2\cdot L$
            \State $m \gets 1$
        \EndIf
        \LComment{amplitude amplification}
        \State $k \gets \mathrm{random \;integer \; in \;[0,m)}$
        \State $\ket{\psi} \gets \hspace{-3pt}\sum\limits_{\Vec{a} \in \mathcal{A}^{\otimes L }}\hspace{-5pt} \sqrt{\pi(\Vec{a})} \ket{\Vec{a}}_A \ket{\vec{s}_\mathrm{init}}_S \ket{-}_R$   
        \State $\ket{\psi'} \gets G^k \ket{\psi}$ 
        \State $\Vec{a}' \gets$ \textbf{measure} $\ket{\psi'}$          
        \State $N_\mathrm{act} \gets N_\mathrm{act} + 2\cdot k\cdot L$
        \LComment{classical verification episode}
        \For{$i=1$ \textbf{to} $L$}
            \State $s_i \gets S_i(s_{i-1}, a_i')$
            \State $r_i \gets R_i(s_{i-1}, a_i')$
            \State $N_\mathrm{act} \gets N_\mathrm{act} + 1$
            \If{$r_i=1$}
                \State \Return $\Vec{a}'$, $N_\mathrm{act}$, $(s_0, \dots, s_i)$, $r_i$
            \EndIf
        \EndFor
        \LComment{no reward}
        \State $m \gets \min\left( \lambda \cdot m, \sqrt{|\mathcal{A}|^L} \right)$ 
    \EndWhile
\end{algorithmic}
\end{algorithm}

Our learning algorithm (see \Cref{alg:prob-jump}), is strongly intertwined with this notion of two stages in Boyer's algorithm.
The algorithm starts with a minimal episode length, which is $L=1$ in the most uninformed case.
We set the lower bound estimate $p_\mathrm{min} = | \mathcal{A} |^{-L}$, assuming an uniform initial policy and at least one rewarded action sequence at the current episode length.
In each round of the main loop, the episode length has a chance to be doubled, with probability 
\begin{align}\label{eq:jump-probability}
    \hspace{-5pt} \varphi_L(m) \equiv \frac{\log(m)}{\log \left( \sqrt{ |\mathcal{A}|^L } \right) } = \frac{2\log(m)}{L \log(| \mathcal{A} |)}.
\end{align}
This probability is chosen such that it reaches unity exactly when Boyer's algorithm reaches its critical stage.
If the episode length doubling is triggered, $m$ is reset to one.

This probabilistic episode length selection strategy serves several purposes.
First, starting from low episode length values is resource-friendly, since the Hilbert space of action sequences, $\mathcal{H}_A^{\otimes L}$, scales exponentially with the episode length $L$.
It is also more efficient with regard to the total number of actions played before reaching the target, which will be our main metric (see \Cref{sec:results:sub:figure_of_merit}).
Second, the exponential increase of $L$ enables the hybrid algorithm to reach large episode lengths reasonably quickly, which might be necessary in scenarios with distant targets.
Finally, coupling the doubling probability to the parameter $m$ ensures that the algorithm does not spend too many tries with an episode length which has no, or vanishing, success chance.

\subsection{Simulation}
\label{sec:methods:sub:simulation}

\begin{figure}[!b]
    \centering
    \includegraphics[width=\columnwidth]{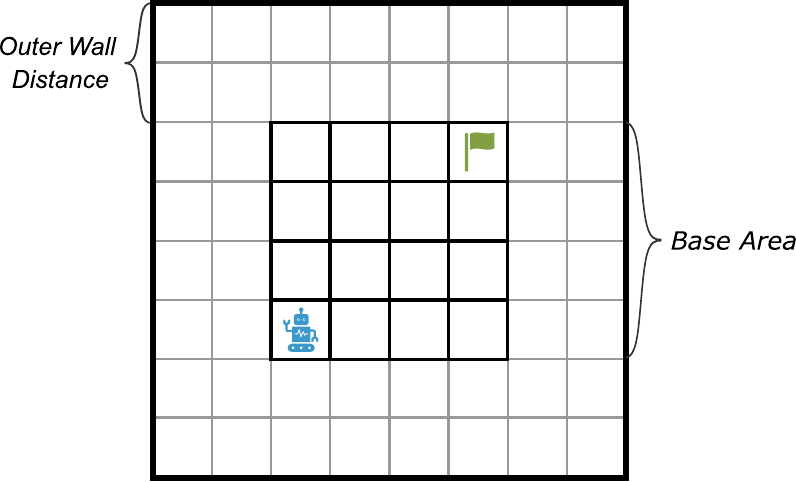}
    \caption{Example of a Gridworld layout used in the simulations. 
    The blue robot and the green flag symbols indicate the start and target position, respectively. 
    The inner square of thick lines is the so-called \textit{base area}, here with a size of $4\times 4$ cells. 
    The Gridworld has outer walls which prohibit the RL agent from moving away further. 
    The example here shows an \textit{outer wall distance} of 2.}
    \label{fig:gridworld_testing_layout}
\end{figure}
To systematically test our hybrid method, we focus on finding the first reward in a two-dimensional Gridworld scenario.
The Gridworld layouts in our simulations have a quadratic \textit{base area} without inner walls with one start and one target cell placed in diagonally opposing corners.
In each cell, the RL agent may choose one action of the set $\mathcal{A}\in \{\text{up, down, left, right}\}$.
The first action to reach the target cell in each episode yields a reward of one; every other action is not rewarded.
Further, the quadratic base area is surrounded by outer walls.
Standing next to a wall and choosing an action towards it yields no change in the cell state.
\begin{figure*}[!t]
    \centering
    \includegraphics[width=0.7\textwidth]{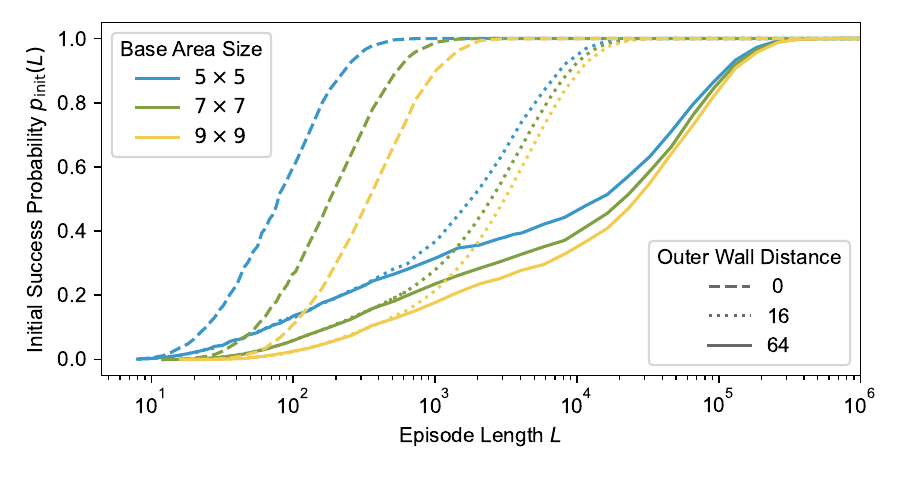}
    \caption{Dependency of the agent's initial success probability on its episode length $L$ for different Gridworld layouts (varied by their base area size and outer wall distance). 
    The probability values are generated using a Monte Carlo simulation, see Appendix~\ref{app:simulation_details} for more details.}
    \label{fig:p_init_overview}
\end{figure*}
Given that in the Gridworld scenario, intermediate percepts of an episode depend on the actions played within that episode, the hybrid algorithm requires $\alpha_k=2$ queries of the environment unitary~$U_\mathrm{env}$ per iteration of the Grover operator~$G$ (cf.~\cref{ftn:alpha-k} on \cpageref{ftn:alpha-k}).

To generate a variety of shapes for the function $p_\mathrm{init}(L)$  on which the performance of our strategies depends, we vary this basic Gridworld layout in two ways, as illustrated in Fig.~\ref{fig:gridworld_testing_layout}:
\begin{enumerate}
    \item We vary the side lengths of the quadratic base area, which we denote~$b$. 
    This obviously has an effect on the \textit{minimal episode length} $L_\mathrm{min}$ necessary to reach the target, $L_\mathrm{min}=2(b-1)$. 
    Additionally, it influences the initial success probability for the minimal episode length (assuming a uniform policy initialization): $p_\mathrm{init}(L_\mathrm{min})=\binom{2(b-1)}{b-1}\cdot 4^{-2(b-1)}$, with $\binom{2(b-1)}{b-1}$ being the number of distinct paths of length $2(b-1) = L_\mathrm{min}$ that reach the target.
    \item We vary the distance of the outer walls around the Gridworld's base area.
    An \textit{outer wall distance}, or $d_\mathrm{wall}$, of zero indicates that the walls are directly surrounding the base area.
    Having $d_\mathrm{wall} = n$ would result in a ring of cells of thickness $n$ between the base area and the outer walls.
    This has no effect on $p_\mathrm{init}(L_\mathrm{min})$. However, more distanced outer walls increase the general state space and, in particular, add cells to the state space which are further from the target cell than any cell in the Gridworld's base area.
    Therefore, it effectively decreases how quickly $p_\mathrm{init}(L)$ rises with increasing episode length $L$.
\end{enumerate}

In our simulations, we vary the Gridworld's base area from size $5 \times 5$ (with $p_\mathrm{init}(L_\mathrm{min})=\num{1.1e-03}$) to $9 \times 9$ ($p_\mathrm{init}(L_\mathrm{min})=\num{2.9e-06}$).
For the outer wall distance, we test the values \numlist[list-final-separator = {, and }]{0;4;8;16;32;64}.
The scenario of no outer walls, which is equal to the limit of an infinite outer wall distance, is not computationally feasible to realize within our simulation framework (for an explanation and more implementation details, see Appendix~\ref{app:simulation_details}.)
The dependency of $p_\mathrm{init}(L)$ on a selection of the different Gridworld layouts is shown in~Fig.~\ref{fig:p_init_overview}.

Since we focus on the scenario of finding the first reward, we assume an untrained agent with an initial uniform policy $\pi(a) = \frac{1}{|\mathcal{A}|} \quad \forall \, a \in \mathcal{A}$.
Every strategy is tested on each Gridworld layout for $N=\num{10000}$ runs.

\subsection{Classical Strategies}
\label{sec:methods:sub:classical-strategies}

We compare the extended hybrid algorithm to two classical strategies, which we present and motivate in the following.

A direct classical equivalent to the extended hybrid algorithm can be devised by employing the same probabilistic episode length doubling strategy.
Here, the parameter $m$ of \Cref{alg:prob-jump} loses its twofold function and only defines the respective probability to double the episode length $L$.
This episode length then defines the number of steps the agent may take until it is reset to its starting position.
From here on, we denote this the \textit{probabilistic classical} strategy.

The second classical strategy arises from the idea that only in the hybrid algorithm an episode length needs to be given.
Classical algorithms, however, are not restricted in such a way.
Practically, not setting an episode length implies an uninterrupted random walk governed by the agent's current policy until the reward state is reached.
We denote this the \textit{unrestricted classical} strategy.

\section{Results}
\label{sec:results}

Before presenting the results of our simulation, we discuss the novel figure of merit and its implications. 
This figure of merit is different from the original hybrid learning agent introduced in~\citep{hamann2022performance}.

\begin{figure*}[!ht]
    \centering
    \includegraphics[width=0.88\textwidth]{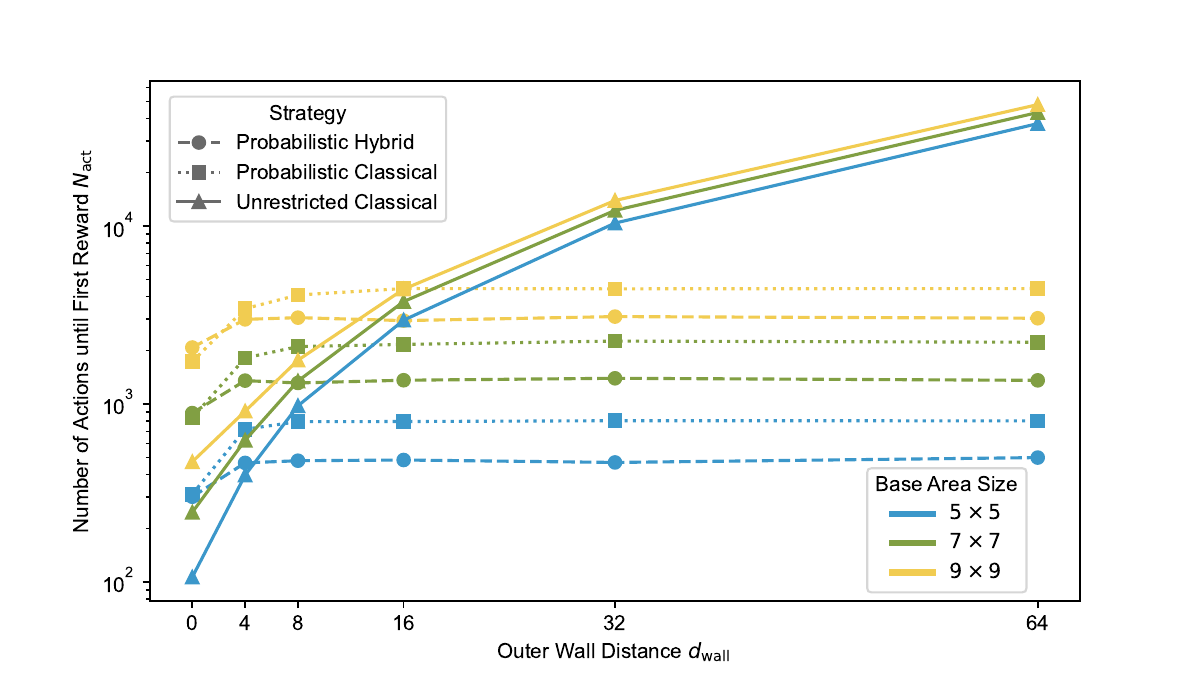}
    \caption{Results of the tested hybrid and classical strategies for the first reward search problem. 
    Gridworld layouts are varied by their base area size as well as their outer wall distance.}
    \label{fig:results}
\end{figure*}

\subsection{Figure of Merit}
\label{sec:results:sub:figure_of_merit}

The quadratic speedup that was proven theoretically and experimentally in the initial works on the hybrid learning agent~\citep{hamann2022performance,saggio2021experimental} is based on the number of queries of the environment unitary $U_\mathrm{env}$.
This is equivalent to the number of episodes played in a classical context.

This figure of merit is misleading in this scenario, as we now argue.
Unlike in~\citep{hamann2022performance}, we no longer operate with a fixed episode length in our RL scenario.
Given that the episode length has a crucial impact on the agent's initial success probability (cf. Fig.~\ref{fig:p_init_overview}), omitting an episode's length from the figure of merit is unreasonable in this scenario.
Additionally, due to the monotonously increasing success probability, in the limit of an infinite episode length one singular episode is always sufficient to find the target.

Hence, we define the total number of actions taken, $N_\mathrm{act}$, instead of the number of RL episodes as the new figure of merit in our scenario.
With a current episode length of $L$, this metric is counted as follows:
\begin{itemize}
    \item In the quantum part of the hybrid learning agent, $k$ iterations of amplitude amplification add $2\cdot k \cdot L$ steps to the count.
    Here, the factor two stems from the fact that we require $\alpha_k=2$ applications of the environment unitary~$U_\mathrm{env}$ per iteration of amplitude amplification.
    \item In the classical part of the hybrid learning agent and for a purely classical agent, an unsuccessful episode adds $L$ steps to the count. If the agent reaches the target, only the actual number of steps $i \leq L$ necessary to reach the target is counted.
\end{itemize}

\subsection{Simulation Results}
\label{sec:results:sub:results}

A full overview of the results for each combination of strategy and Gridworld configuration is given in \Cref{tab:full_results} of Appendix~\ref{app:full_results}.
Fig.~\ref{fig:results} visualizes the results on a subset of Gridworld configurations.

According to the two ways by which we varied the basic Gridworld layout, \textit{base area size} $b$ and \textit{outer wall distance} $d_\mathrm{wall}$, two effects can be observed in the data: 
\begin{itemize}
    \item[(i)] The necessary number of actions to reach the target increases with increasing base area size $b$ as expected. 
    This observation holds for all strategies and outer wall distance values. 
    \item[(ii)] The influence of the outer wall distance parameter, $d_\mathrm{wall}$, differs for the different strategies. 
    Thus, the question whether the hybrid agent or one of the classical agents is preferable depends on the outer wall distance.
\end{itemize}

At $d_\mathrm{wall}=0$ and $d_\mathrm{wall}=4$, the unrestricted classical agent requires on average the least number of actions to reach the target across all base area sizes.
Both probabilistic strategies require approximately \SIrange{1.2}{5}{} times the number of actions, with the hybrid version requiring the most steps at larger base area sizes.
For larger values of $d_\mathrm{wall}$, both probabilistic strategies appear to stabilize in terms of $N_\mathrm{act}$, which can be seen from the flat curves in Fig.~\ref{fig:results} for outer wall distances of 8 and higher.
Between the two, the hybrid strategy consistently has lower $N_\mathrm{act}$, with the ratio of fewer actions ranging from \SIrange{27}{42}{\percent} (for base area $9 \times 9$ with $d_\mathrm{wall}=8$ and base area $5\times 5$ and $d_\mathrm{wall}=64$, respectively).
For the unrestricted classical strategy, however, the number of actions increases continuously with increasing $d_\mathrm{wall}$.
At $d_\mathrm{wall}=8$, it is still more efficient than the hybrid agent for the two largest base area sizes, $8\times 8$ and $9 \times 9$.
Already for $d_\mathrm{wall}=16$, however, it requires more actions than either probabilistic strategy for any base area size.
For the largest tested Gridworld configuration (base area size $9 \times 9$ and $d_\mathrm{wall}=64$), the unrestricted classical strategy trails the probabilistic hybrid one by more than an order of magnitude.

\begin{figure}[!b]
    \centering
    \includegraphics[width=\columnwidth]{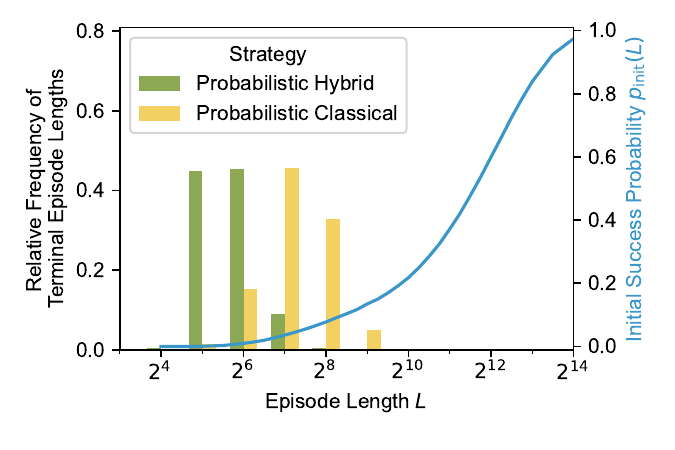}
    \caption{Relative frequency of terminal episode lengths for the probabilistic hybrid and classical algorithms. 
    The blue curve represents the initial success probability for the given Gridworld layout depending on the episode length.
    The results shown stem from the Gridworld configuration with a base area size of $9 \times 9$ and and outer wall distance of 16.}
    \label{fig:final_ep_length_histogram}
\end{figure}

\begin{figure}[!ht]
    \centering
    \includegraphics[width=\columnwidth]{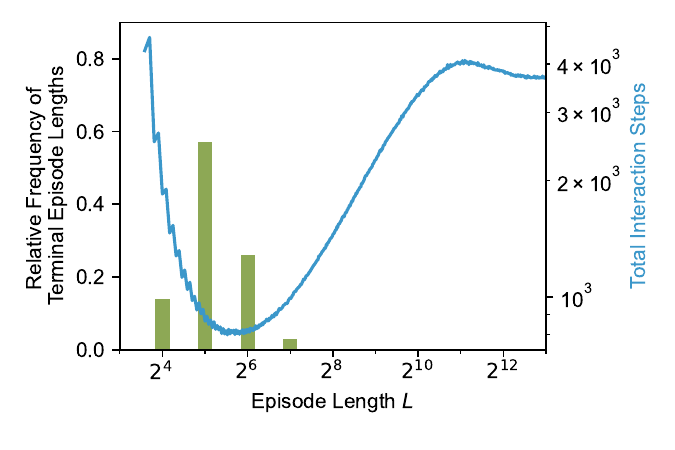}
    \caption{Relative frequency of terminal episode lengths for the probabilistic hybrid algorithm and expected number of total interaction steps for the fixed-length hybrid algorithm (cf.~Fig.~\ref{fig:expected_cost_fix_length}). The Gridworld configuration used here has a base area size of $7\times 7$ and an outer wall distance of 16.}
    \label{fig:terminal_ep_length_plus_expected_cost}
\end{figure}

Another interesting quantity is the episode length at which the strategies terminate, coined \textit{terminal episode length}.
For the probabilistic strategies, this value is always a power of two due to the episode length doubling nature of the algorithms.
A visual comparison of the relative frequencies with which either probabilistic strategy terminates at a specific episode length is given in Fig.~\ref{fig:final_ep_length_histogram}.
The probabilistic hybrid strategy terminates on average at lower episode lengths than the classical probabilistic strategy.
For the Gridworld configuration used for~Fig.~\ref{fig:final_ep_length_histogram} (base area $9\times 9$ and $d_\mathrm{wall}=16$), the most frequent terminal episode lengths for the hybrid strategy are $2^5=32$ and $2^6=64$, whereas the probabilistic classical strategy terminates most frequently at $2^7=128$ and $2^8=256$.
In~Fig.~\ref{fig:terminal_ep_length_plus_expected_cost}, we compare the relative frequency of terminal episode lengths of the probabilistic hybrid algorithm with the total number of interaction steps that the original hybrid learning agent of~\citep{hamann2022performance} would require with a fixed episode length (cf.~Fig.~\ref{fig:expected_cost_fix_length}).

\begin{figure*}[!htb]
    \centering
    \begin{subfigure}[c]{0.49\textwidth}
    \includegraphics[width=\linewidth]{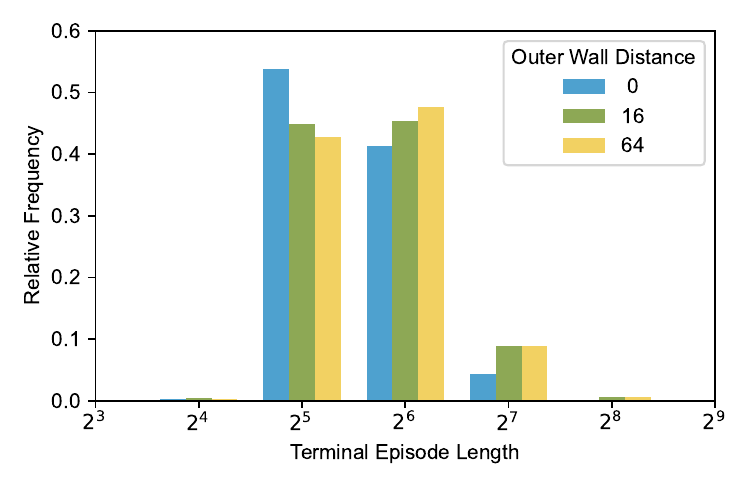}
    \caption{Probabilistic hybrid strategy.}
    \label{fig:histograms:sub:probabilistic_hybrid}
    \end{subfigure}
    \begin{subfigure}[c]{0.49\textwidth}
    \includegraphics[width=\linewidth]{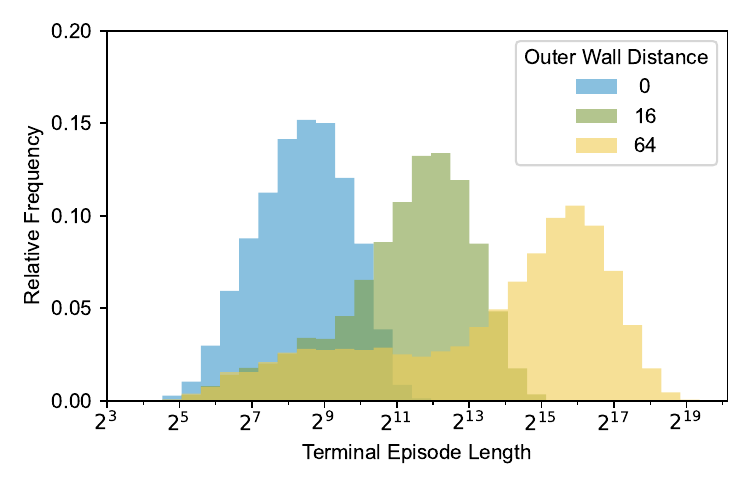}
    \caption{Unrestricted classical strategy.}
    \label{fig:histograms:sub:unrestricted_classical}
    \end{subfigure}
    \caption{Comparison of the terminal episode lengths between the probabilistic hybrid and the unrestricted classical strategy. The results shown are for the configuration with a base area size of $9 \times 9$.}
    \label{fig:histograms}
\end{figure*}

For the given Gridworld configuration, using the hybrid learning agent with fixed episode length $L=34$ would be optimal in terms of the total interaction steps.
As the distribution of terminal episode lengths in~Fig.~\ref{fig:terminal_ep_length_plus_expected_cost} shows, our probabilistic hybrid algorithm terminates most frequently at the nearest ($L=2^5=32$) and second most frequently at the second nearest ($L=2^6=64$) episode length.
This suggests that the in-built episode length doubling mechanism is tuned well enough that the probabilistic hybrid algorithm reaches efficient episode lengths and at the same time does not massively overshoot to unnecessary large episode lengths.

Fig.~\ref{fig:histograms} shows the influence of the outer wall distance parameter on the terminal episode length.
For the probabilistic hybrid strategy (Fig.~\ref{fig:histograms:sub:probabilistic_hybrid}), increasing $d_\mathrm{wall}$ results in a slight shift of relative frequencies towards larger terminal episode lengths.
Namely, the most frequent terminal episode length shifts from 32 for $d_\mathrm{wall}=0$ to 64 for $d_\mathrm{wall}=16$ and $d_\mathrm{wall}=64$.
For the unrestricted classical strategy (Fig.~\ref{fig:histograms:sub:unrestricted_classical}), the episode length shifts by several orders of magnitude for larger outer wall distances.
As the figure shows, for $d_\mathrm{wall}=0$, it terminates most frequently at episode lengths between $2^8$ and $2^9$, whereas for $d_\mathrm{wall}=16$ this interval ranges from $2^{12}$ to $2^{13}$ and for $d_\mathrm{wall}=64$ from $2^{15}$ to $2^{16}$.

\section{Conclusion}
\label{sec:conclusion}
In this work, we have introduced a hybrid agent for quantum-accessible reinforcement learning with a flexible episode length selection strategy. 
As we have argued, this extension is crucial for finding the first reward in RL scenarios which are equivalent to a Gridworld with no information about the length of the shortest rewarded path.

The simulation results for different Gridworld configurations suggest that our proposed hybrid agent can (i) find the first reward faster and (ii) can find shorter solutions than the considered classical agents in many configurations.

From the two options to vary the Gridworld layout, the outer wall distance proves to strongly influence which agent performs best.
Whereas the unrestricted classical strategy requires more and more steps with increasing outer wall distance, the cost of the probabilistic strategies virtually remains the same after $d_\mathrm{wall}\geq 8$.
In the unrestricted classical case, the restricted state space from nearby walls helps the agent, since it can effectively never distance itself further from the target than it is from the starting position. 
With an increasing outer wall distance, however, the agent may very well initially move in a wrong direction and thus effectively increase its distance from the target.
Here, resetting the agent to its starting position is beneficial, which is exactly what the probabilistic strategies do due to their finite episode length.
Given that the probabilistic strategies terminate at much lower episode lengths than the unrestricted classical strategy (cf. Figs.~\ref{fig:final_ep_length_histogram} and \ref{fig:histograms}), they are hardly affected by the ever increasing state space, as the flat curves in Fig.~\ref{fig:results} indicate.

Between the two probabilistic strategies, the hybrid one proves to find the target quicker in all layouts except the most trivial one with  $d_\mathrm{wall}=0$.
Whereas the quadratic scaling advantage does not apply any more for this metric and scenario, the hybrid strategy outperforms its classical counterpart in many cases.
Especially towards lower initial success probabilities and larger state spaces, i.e., harder problems, the hybrid agent's amplitude amplification appears to be advantageous.

The effect of increasing the Gridworld's base area size is the same for all agents: it makes the problem harder overall.
Comparing the distributions of terminal episode lengths of the different strategies and  the initial success probability curve $p_\mathrm{init}(L)$ in Fig.~\ref{fig:final_ep_length_histogram} also reveals that the doubling of the episode length happens slow enough such that (i) the maximal episode length stays in a reasonable regime and that (ii) noticeable speedups through amplitude amplification can be achieved. 
In addition, the lower terminal episode length suggests that the solutions found by the hybrid agent are shorter on average than the ones found by the probabilistic classical strategy.
This can lead to additional advantages in the further learning resulting in, e.g., faster or better learning results. 
Further, the comparison with our motivating example in~Fig.~\ref{fig:terminal_ep_length_plus_expected_cost} suggests that even without prior knowledge of the optimal episode length, the probabilistic hybrid algorithm terminates most frequently at the episode lengths closest to it.

Finally, we address a few design questions on the chosen Gridworld layout, especially regarding a few omissions of further Gridworld variations.
First, one could conceive of a scenario where moving into some cells, or even all walls, stops the episode immediately without a reward.
If we have a move sequence $\Vec{a}$ which moves into such a terminal, but non-rewarded cell, concatenating any additional move sequence $\Vec{a}'$ will not turn the full sequence into a rewarded one.
Thus, with our initial uniform policy, the initial success probability does not converge to 1 in the limit of infinite episode lengths.
Here, it can be assumed that the probabilistic hybrid strategy is beneficial as this strategy works well with a slowly increasing success probability and low probabilities in general.
Second, one could omit the outer walls altogether, yielding an infinite state space.
As mentioned in \Cref{sec:methods:sub:simulation}, simulating this scenario is not feasible.
However, we can extrapolate the trend for increasing outer wall distances, since having no outer walls is equal to the limit  $d_\mathrm{wall}\rightarrow \infty$. 
Here, we can see that the probabilistic strategies prove to be advantageous, with the hybrid version still requiring less steps than the classical one.
Third and last, one could investigate higher-dimensional Gridworld layouts than the two-dimensional scenario shown here.
For random walks in hypergrids of dimension $D\geq 3$, however, the probability to reach any point with a random walk in the limit of infinite steps does not converge to unity~\citep{polya1921aufgabe}.
Therefore, an infinite random walk in the fashion of the unrestricted classical strategy is certainly a bad choice.
Given that, besides the asymptotic limit, the scenario is not fundamentally different, the hybrid probabilistic strategy can again be assumed to be the most efficient of the three, supposed that the initial success probability is low.

There are two research questions which should be investigated further: 
(i) Is the coincidence of the most probable  terminal episode length with the optimal episode length pure luck or a reliable property of our algorithm? 
(ii) In which ways does the on average shorter terminal episode length of our probabilistic hybrid agent compared to the probabilistic classical agent influence the further learning? 
In addition, the successful extension of the hybrid learning agent to episodic learning tasks with unknown target distance now enables the application to many more realistic learning tasks, which should be investigated in the future.

\backmatter

\bmhead{Acknowledgements}
This project was made possible by the DLR Quantum Computing Initiative and the Federal Ministry for Economic Affairs and Climate Action; qci.dlr.de/projects/qlearning.
The authors gratefully acknowledge the scientific support and HPC resources provided by the German Aerospace Center (DLR). 
The HPC system CARA is partially funded by \textquote{Saxon State Ministry for Economic Affairs, Labour and Transport} and \textquote{Federal Ministry for Economic Affairs and Climate Action}.
The authors thank Alessio Belenchia and Benjamin Desef for helpful comments.

\section*{Declarations}

\bmhead{Competing interests}
The authors declare no competing interests.


\begin{appendices}

\begin{landscape}
\section{Table of Full Results}
\label{app:full_results}

\begin{table}[!hb]
    \caption{Results of the tested hybrid and classical strategies for the first reward search problem. 
    Numbers represent average $N_\mathrm{act}$ (with the respective standard error in parentheses) for the different Gridworld layouts, based on $N=\num{10000}$ runs per configuration. 
    For each configuration, the lowest value of $N_\mathrm{act}$ is printed in boldface.}
    \label{tab:full_results}
    \begin{tabular}{@{} S[table-format=2.0] l @{\hspace{2em}} *5{S[detect-weight,mode=text,table-format=5.5]}}
        \toprule
        {\multirow[c]{2}{3cm}[-3pt]{Outer Wall Distance}} & {\multirow[c]{2}{3.2cm}[-3pt]{Strategy}} & \multicolumn{5}{c}{Base Area Size} \\ \cmidrule[0.3pt]{3-7}
         & & {$5 \times 5$} & {$6 \times 6$} & {$7 \times 7$} & {$8 \times 8$} & {$9 \times 9$} \\ [-1pt]
        \midrule
         & Probabilistic Hybrid & 300 \pm 2 & 515 \pm 3 & 887 \pm 5 & 1400 \pm 9 & 2071 \pm 14 \\
        0 & Probabilistic Classical & 310 \pm 2 & 534 \pm 3 & 833 \pm 5 & 1208 \pm 7 & 1724 \pm 9 \\
         & Unrestricted Classical & \B 106 \pm 1  & \B 170 \pm 2 & \B 246 \pm 2 & \B 347 \pm 3 & \B 472 \pm 4 \\ 
        \cmidrule(l{.75em}r{.25em}){1-7}
         & Probabilistic Hybrid & 466 \pm 3 & 806 \pm 4 & 1350 \pm 7 & 2087 \pm 11 & 2981 \pm 17 \\
        4 & Probabilistic Classical & 720 \pm 6 & 1206 \pm 9 & 1809 \pm 12 & 2541 \pm 16 & 3426 \pm 21 \\
         & Unrestricted Classical & \B 397 \pm 4 & \B 503 \pm 5 & \B 624 \pm 6 & \B 749 \pm 7 & \B 909 \pm 8 \\ 
        \cmidrule(l{.75em}r{.25em}){1-7}
         & Probabilistic Hybrid & \B 479 \pm 3 & \B 797 \pm 5 & \B 1309 \pm 8 &  2106 \pm 14 &  3048 \pm 18 \\
        8 & Probabilistic Classical & 793 \pm 7 & 1333 \pm 11 & 2096 \pm 16 & 3083 \pm 23 & 4085 \pm 29 \\ 
         & Unrestricted Classical & 975 \pm 10 & 1165 \pm 11 & 1343 \pm 13 & \B 1539 \pm 14 & \B 1758 \pm 16 \\ 
        \cmidrule(l{.75em}r{.25em}){1-7}
         & Probabilistic Hybrid & \B 483 \pm 3 & \B 826 \pm 4 & \B 1357 \pm 7 & \B 1986 \pm 11 & \B 2934 \pm 17 \\
        16 & Probabilistic Classical & 795 \pm 7 & 1374 \pm 12 & 2157 \pm 18 & 3115 \pm 24 & 4442 \pm 34 \\ 
         & Unrestricted Classical & 2950 \pm 34 & 3375 \pm 37 & 3754 \pm 41 & 4095 \pm 43 & 4406 \pm 45 \\ 
         \cmidrule(l{.75em}r{.25em}){1-7}
         & Probabilistic Hybrid & \B 468 \pm 3 & \B 816 \pm 5 & \B 1391 \pm 8 & \B 2047 \pm 12 & \B 3090 \pm 18 \\
        32 & Probabilistic Classical & 803 \pm 7 & 1364 \pm 12 & 2250 \pm 18 & 3291 \pm 26 & 4426 \pm 34 \\ 
         & Unrestricted Classical & 10357 \pm 135 & 11219 \pm 139 & 12208 \pm 151 & 13152 \pm 154 & 13872 \pm 157 \\ 
         \cmidrule(l{.75em}r{.25em}){1-7}
         &   Probabilistic Hybrid & \B 499 \pm 3 & \B 847 \pm 5 & \B 1355 \pm 7 & \B 2173 \pm 12 & \B 3025 \pm 18 \\ 
        64 & Probabilistic Classical & 802 \pm 7 & 1385 \pm 12 & 2218 \pm 18 & 3147 \pm 25 & 4442 \pm 34 \\ 
         & Unrestricted Classical & 37514 \pm 543 & 40668 \pm 555 & 43315 \pm 578 & 46448 \pm 606 & 47963 \pm 615 \\ 
         \bottomrule
    \end{tabular}
\end{table}
\end{landscape}

\section{Simulation Details}
\label{app:simulation_details}

Given that amplitude amplification (AA) is not a NISQ-compatible algorithm, we have to fall back to simulating its effect instead of doing a full quantum circuit execution on simulated or real hardware.

To do so, we  infer the initial success probability $p_\mathrm{init}(L)$ for any episode length $L$ which might occur in the hybrid strategy beforehand.
Having knowledge of this probability, we can subsequently compute the amplified success probability using the well-known AA equation~\citep{brassard2002quantum}
\begin{align*}
    p_{AA}(L,k) = \sin^2\left(\left[2k+1\right]\arcsin\left[p_\mathrm{init}(L)^{-1/2}\right]\right)
\end{align*}
for $k$ iterations of our Grover operator $G$.
This probability can in turn be used to correctly sample a rewarded or non-rewarded action sequence.

Given that the initial policy which generates $p_\mathrm{init}(L)$ is a uniform probability distribution over the space of actions, the agent's movement initially equals an unweighted random walk.
Therefore, we can estimate $p_\mathrm{init}(L)$ with a Monte Carlo simulation of random walks of length $L$.

For each combination of Gridworld configuration and episode length $L$ to be tested, we perform at least $N_\mathrm{shots}(L)=2^{14}$ runs and count the number of random walks which terminated successfully, $N_\mathrm{success}(L)$ (i.e., which have the target cell in their path).
To improve numerical stability, we keep incrementing $N_\mathrm{shots}(L)$ in batches of $2^{14}$ until $N_\mathrm{success}(L)$ has reached at least 16.
Doing so, we can finally estimate the initial success probability simply as the ratio of successes to shots:
\begin{align*}
    p_\mathrm{init}(L) \approx \frac{N_\mathrm{success}(L)}{N_\mathrm{shots}(L)}.    
\end{align*}

Due to the hybrid agent's doubling strategy, $p_\mathrm{init}(L)$ only needs to be pre-computed for powers of 2 and until convergence of $p_\mathrm{init}(L)$.
For the plot in Fig.~\ref{fig:p_init_overview}, however, we also computed $p_\mathrm{init}(L)$ for intermediate values to create a smoother curve using linear interpolation.

Finally, in this section we address why omitting outer walls at all is not feasible within this framework.
As proven in~\citep{polya1921aufgabe}, on an infinite two-dimensional grid, the random walker's probability to pass by any given point $\mathbf{x}\in \mathbb{Z}^2$ converges to 1 in the limit of infinite steps.
Therefore, even for the scenario of no walls, which is equivalent to an infinite two-dimensional grid, $p_\mathrm{init}(L)$ should converge towards one in the limit of infinite steps, $$\lim_{L \rightarrow \infty} p_\mathrm{init}(L) = 1.$$

The issue for our simulation is, however, the slow rate of convergence.
Even for the smallest base area size of $5\times 5$, $p_\mathrm{init}(L)$ has just reached approximately \SI{60}{\percent} for $L=2^{22}=\num{4194304}$ in the ``no-walls'' scenario, whereas for $d_\mathrm{wall}=64$ the probability already converges near $L=2^{19}=\num{1048576}$.
By counting just the steps of unsuccessful random walks, we thus arrive at 
\begin{align*}
    & L \cdot (1 - p_\mathrm{init}(L)) \cdot N_\mathrm{shots}\\
    \approx&\, 2^{22}\cdot (1-0.6) \cdot 2^{14}\\
    \approx&\, \num{2.7e10}
\end{align*}
steps computed just for this episode length, which becomes soon fully intractable for even larger episode lengths due to the slow increase in $p_\mathrm{init}(L)$.

Additionally, the run time scaling of the unrestricted classical strategy in~Fig.~\ref{fig:results} with increasing outer wall distance shows the intractability of simulating this strategy in a ``no-walls'' scenario.

\section{Total Interaction Steps for Fixed Episode Length}
\label{app:fixed-episode-length}
In this section, we give some background on the performance comparison shown in~Fig.~\ref{fig:expected_cost_fix_length}, which motivates the flexible episode length selection strategy for the hybrid agent.

The example stems from an RL setting of a Gridworld with a base area size of $7 \times 7$ and an outer wall distance of~16 (see \Cref{sec:methods:sub:simulation} for our Gridworld layout definitions).
We choose fixed episode lengths $L$ in the interval ranging from $L_\mathrm{min}=12$ up to $2^{14}=\num{16834}$.
Further, we assume untrained agents initialized with uniform action selection probabilities such that the classical success probability $p_\mathrm{init}(L)$ is the one of a random walk of length $L$ through the maze.
For both the classical and hybrid agent, we count the total number of interaction steps, i.e., the total number of actions performed until the first reward is reached.

The results presented in~Fig.~\ref{fig:expected_cost_fix_length} are generated with a Monte Carlo simulation of \num{100000} repetitions each on a logarithmically-spaced grid of \num{868} different episode length values.
For the classical agent, we perform a random walk that is periodically reset after $L$ steps until the first reward is found, aggregating the total number of steps.
For the hybrid agent, we rely on the simulation of amplitude amplification presented in Appendix~\ref{app:simulation_details}, using the precomputed initial success probabilities for each episode length.
If the simulated amplitude amplification returns a rewarded episode, we sample the length of a rewarded action sequence by performing random walks of at most length $L$ until there is a rewarded one.

The ``zig-zag'' behavior in~Fig.~\ref{fig:expected_cost_fix_length} for small episode lengths can be explained as follows.
For the chosen Gridworld layout, the minimal episode length to reach the target is twelve. 
As we increase the episode length, non-optimal paths may now also reach the target, resulting in an increase of the success probability.
However, this increase only occurs in episode length intervals of two as one cannot land on the target cell with an odd number of steps.
Thus, for small episode lengths, the success probability only increases from an even number to the next but stays constant for the next larger odd value. 
Only when the episode length is large enough that the agent may run into a wall and thus ``waste'' a step, the success probability increases for every incrementally larger episode length.
This piecewise constant success probability leads to the spikes for odd episode length values in~Fig.~\ref{fig:expected_cost_fix_length}.
Indeed, there are on average as many unsuccessful episodes as with the next lower even episode length, but these episodes are more ``costly'' due to the additional step.

\end{appendices}

\end{document}